\documentclass{article}
\usepackage{graphicx}

\begin{document}

\title{Chaos in an intermittently driven damped oscillator}
\author{M. P. John and V. M. Nandakumaran\\
International School of Photonics\\ Cochin University of Science and
Technology.\\Cochin 682 022, India\\}

\maketitle
\begin{abstract}
We observe chaotic dynamics in a damped linear oscillator,
which is driven only at certain regions of phase space. Both
deterministic and random drives are studied. The dynamics is
characterized using standard techniques of nonlinear dynamics.
Interchanging roles of determinism and stochasticity is also
considered.
\end{abstract}

\section{Introduction}
It is generally known that continuous linear systems cannot exhibit
chaotic time evolution. However recently it has been shown that
chaotic behavior can occur in systems without explicit non
linearities. Hirata et.al. \cite{pulse} have shown that continuous
chaotic wave forms can be constructed by the superposition of
certain pulse basis functions. In a linear second order filter
driven by randomly polarized pulses, chaotic behavior is observed
when the time series is viewed backwards in time \cite{withuot}.
Further they have shown that by reversing the time series, folded
band chaos similar to R\"{o}ssler attractor can also be synthesized
\cite{folded}. Even if the time reversal is not physical, this
reveals the importance of pure randomness which is associated with
chaotic behavior.  Studies in this direction are important in
understanding the relation between randomness, determinism and
complexity in dynamical systems.

We study a  weakly damped driven linear oscillator where the
oscillator and the drive are coupled only when the trajectory is in
a thin strip in the phase space. This modification introduces strong
nonlinearities required for chaotic behavior. First we consider the
oscillator coupled to a deterministic drive. Then we consider a
random forcing. We show that the attractors formed by both the
deterministic and random forcing are similar in many respects.

\section{Basic model}
The driven damped linear oscillator is well known and is important
in almost all branches of physical science. The dynamics of this
system satisfies the equation,
\begin{eqnarray}
\label{driv_damp} \frac{d^2 x}{d t^2}+ \gamma \frac{d x}{dt}+
\Omega^2 x & = & a\sin(\omega t).
\end{eqnarray}
The parameters in the equation are $\gamma$, the damping
coefficient, $\Omega$, the angular frequency of the oscillator and
$\omega$ the frequency of the drive. Such a system shows periodic
behavior after the initial transients have vanished.

Our basic model is the one in which the system is driven only in a
thin strip of the phase space. We choose the origin as the location
of the strip because  orbits with lower energy may not encounter the
strip, if it is located far from the origin. Such orbits will spiral
towards zero as in the case of an ordinary damped oscillator. When
not driven, the oscillator performs a damped motion and the drive
runs freely. First we consider a deterministic drive. The dynamics
is characterized using standard techniques. Also the behavior under
parameter variations is studied. Later we discuss the effect of
replacing the deterministic driving term with a stochastic one.

\section{Deterministic driving}
\subsection{Dynamical equations}
The equation \ref{driv_damp} modified accordingly by the addition of
a new term $\mu(s,x)$ as,
\begin{eqnarray}
\label{fliker} \frac{d^2 x}{d t^2}+ \gamma \frac{d x}{dt}+ \Omega^2
x & = & \mu(s,x) a\sin(\omega t)
\end{eqnarray}
where,
\begin{equation}\label{strip}
\mu(s,x) = \left\{
\begin{array}{ll}
1 & $if$ ~ \mid x \mid ~ \leq s\\
0 & $otherwise$.\\
\end{array}
\right.
\end{equation}
Thus with a deterministic drive, what the dynamical equations
represent is legitimate deterministic system in every sense.

\subsection{Numerical results} 
\begin{figure}[tbh]
\centering \includegraphics[width=0.8\columnwidth]{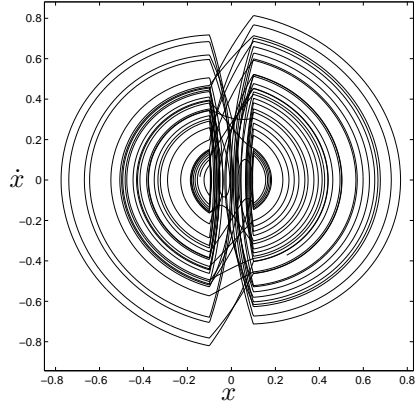}
\caption{Phase space trajectory of the oscillator ($\Omega~=~1.0$,
$\gamma~=~0.1$, $\omega~=~0.3$, $s=0.1$ and $a~=~2.5$).}
\label{phspace}
\end{figure}

\begin{figure}[tbh]
\centering \includegraphics[width=0.8\columnwidth]{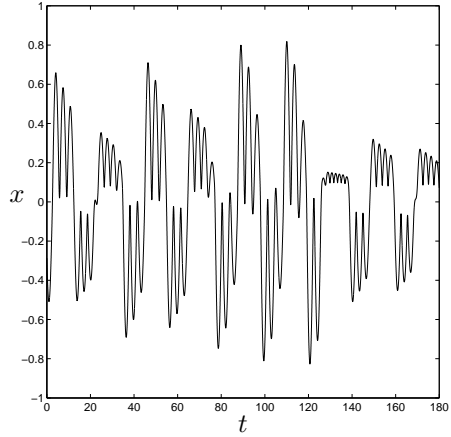}
\caption{Time series of the system ($\Omega~=~1.0$, $\gamma~=~0.1$,
$\omega~=~0.3$, $s=0.1$ and $a~=~2.5$).} \label{series}
\end{figure}

\begin{figure}[tbh]
\centering \includegraphics[width=0.8\columnwidth]{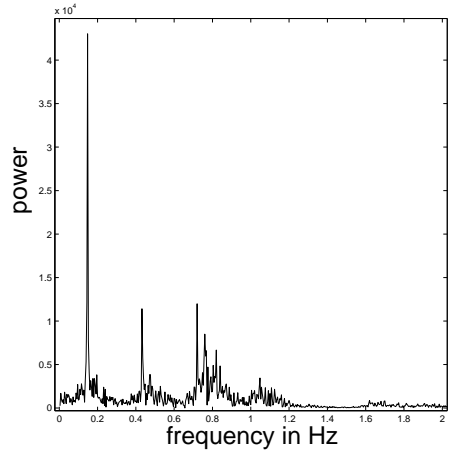}
\caption{Power spectrum of the time series $x(t)$ ($\Omega~=~1.0$,
$\gamma~=~0.1$, $\omega~=~0.3$, $s=0.1$ and $a~=~2.5$).} \label{pow}
\end{figure}

\begin{figure}[tbh]
\centering \includegraphics[width=0.8\columnwidth]{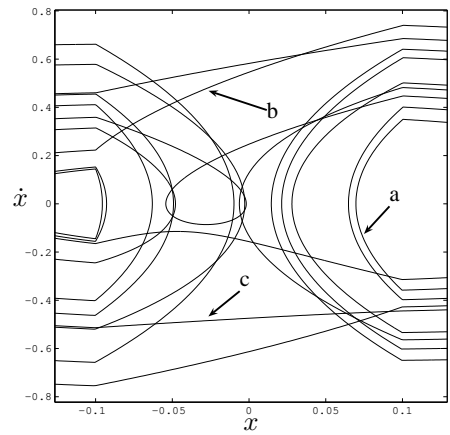}
\caption{Close view of the strip. Note the trajectories: a) reverse
direction b) cross the strip but suffers deformation, c) cross the
strip without considerable modification.} \label{closeup}
\end{figure}

\begin{figure}[tbh]
\centering \includegraphics[width=0.8\columnwidth]{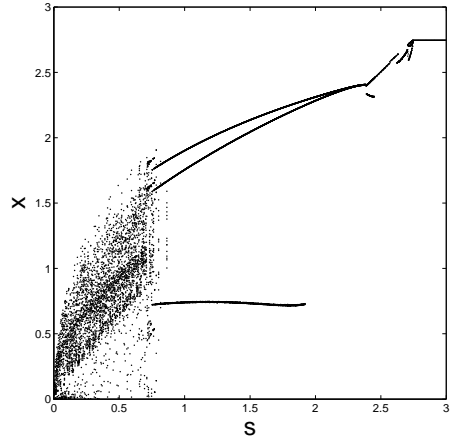}
\caption{Bifurcation diagram obtained by varying the strip width
$s$. Chaotic behavior disappears as the width of the strip is
($\Omega~=~1.0$, $\gamma~=~0.1$, $\omega~=~0.3$ and $a~=~2.5$).}
\label{bifstrip}
\end{figure}

\begin{figure}[tbh]
\centering \includegraphics[width=0.8\columnwidth]{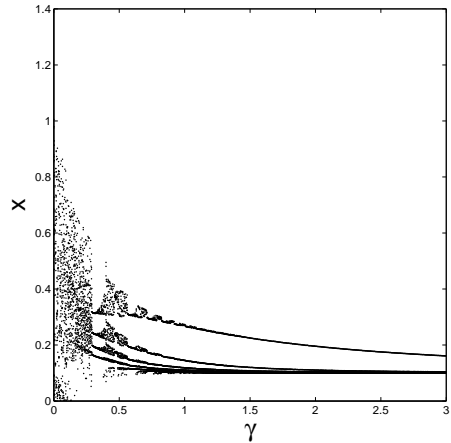}
\caption{Bifurcation diagram obtained by varying the damping
coefficient $\gamma$. Chaotic behavior is suppressed as the damping
is increased ($\Omega~=~1.0$, $\omega~=~0.3$, $s=0.1$ and
$a~=~2.5$).} \label{bifgamma}
\end{figure}

Numerical simulations were done using
the Runge-Kutta algorithm with a step size of $10^{-3}$. We choose
the following values for the parameters:  $\Omega~=~1.0$,
$\gamma~=~0.1$, $\omega~=~0.3$, $s=0.1$ and $a~=~2.5$. The
parameters were chosen by numerical trials to express the concepts
clearly, and are fixed during the evolution of the system in time.
Fig.\ref{phspace} shows the phase space plot of the system. Note
that there is a discontinuity in a strip of width $2s$ near $0$ in
the the phase space, where the oscillator is coupled to the drive. A
closer view of the strip is given in Fig. \ref{closeup}. It can be
seen that the trajectories that approach the origin may either cross
the strip with or without considerable modifications, or it may get
reversed. Fig. \ref{series} shows the time series $x(t)$ of the
system. The amplitude of oscillations, and the number of

oscillations between two consecutive crossings of the strip are
irregular. This corresponds to the phase space dynamics which
consists of reflections in the strip and transitions which modify
the trajectories. The power spectrum as shown in Fig. \ref{pow} is
broad which confirms the aperiodic behavior in time. Varying the
parameters $\gamma$ and $s$, the system exhibits both chaotic and
periodic oscillations. From the bifurcation diagrams shown in Fig.
\ref{bifstrip} and Fig. \ref{bifgamma}, it can be seen that the
chaotic behavior disappears for higher values of $\gamma$ and $s$.
Also note that for $s~\simeq~1$ there is a transition to the case of
an ordinary driven damped linear oscillator, where no chaotic
behavior is expected.

\begin{figure}[tbh]
\centering \includegraphics[width=0.8\columnwidth]{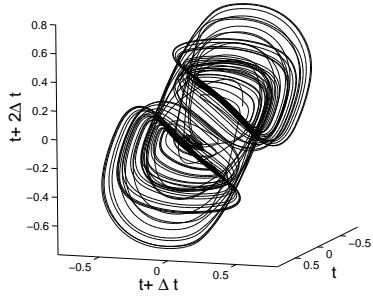}
\caption{The delay reconstructed attractor ($\Omega~=~1.0$,
$\gamma~=~0.1$, $\omega~=~0.3$, $s=0.1$, $a~=~2.5$ and $\Delta
t~=~1.56$ ).} \label{recon}
\end{figure}

We used $DATAPLORE$ \cite{Dataplore} for the analysis of the time
series. The time delay required for reconstructing a timeseries is a
widely discussed topic \cite{addi}. With a delay of 1.56, it is
found that the autocorrelation function falls to $0.5$. With this
delay, embedding dimension was calculated using False nearest
neighbors method \cite{TIS}. It is found that the percentage of
nearest neighbors approaches zero with embedding dimension 3.
Lyapunov Exponents were calculated with embedding dimension 3,
number of nearest neighbors 37 and degree of the extrapolating
polynomial 3. The obtained result is multiplied by the sampling
frequency of the input signal to calculate the Lyapunov exponents.
The Largest Lyapunov exponent is found to be $3.19$. We obtained
2.148 as the Kaplan-Yorke dimension. The reconstructed phase space
of the attractor is shown in fig. \ref{recon} and is similar in
appearance to other chaotic systems.

\section{Random driving}
\subsection{Dynamical equations}
Random forcing is achieved by the drive assuming a random value
$\xi$ each time the phase space trajectory enters the strip. The
amplitude of forcing $\xi$ is Gaussian random variable with variance
$\sigma^2=\frac{1}{\sqrt{2}}$ and zero mean. We define $\chi(s,x)$
as follows,
\begin{equation}\label{xi_strip}
\chi(s,x) = \left\{
\begin{array}{ll}
\xi & $if$ ~ \mid x \mid ~ \leq s\\
0 & $otherwise$.\\
\end{array}
\right.
\end{equation}
Retaining $\mu(s,x)$ defined as in eq. \ref{fliker} for comparison,
the dynamical equation is,
\begin{eqnarray}
\label{fliker_rand} \frac{d^2 x}{d t^2}+ \gamma \frac{d x}{dt}+
\Omega^2 x & = & a \mu(s,x) \chi(s,x)
\end{eqnarray}
\subsection{Numerical results}
\begin{figure}[tbh]
\centering \includegraphics[width=0.8\columnwidth]{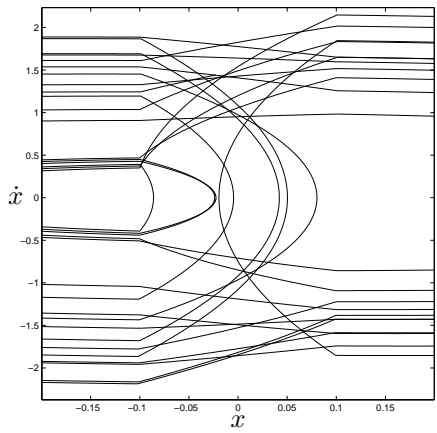}
\caption{Close view of the phase space strip of a randomly driven
oscillator.} \label{strip_xi}
\end{figure}

\begin{figure}[tbh]
\centering \includegraphics[width=0.8\columnwidth]{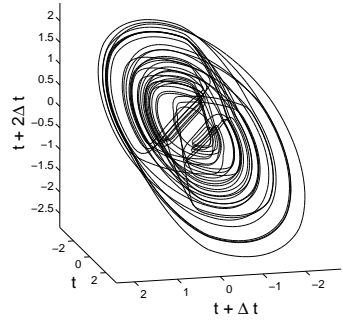}
\caption{The reconstructed attractor resulting from random driving (
$\Omega~=~1.0$, $\gamma~=~0.1$, $s=0.1$, $a~=~2.5$ and $\Delta t ~=
~1.1$).} \label{recon_xi}
\end{figure}

When the oscillator is driven randomly, it is found that the
attractor is still chaotic with the largest Lyapunov exponent equal
to $2.92$ and a Kaplan-Yorke dimension of $2.31$ when embedded in
three dimensions. The dynamics in phase space is similar to the one
obtained with a deterministic drive as shown in Fig. \ref{strip_xi}.
The reconstructed attractor (with time delay $\Delta t ~=~1.1$)
given in Fig. \ref{recon_xi} is similar to the one obtained with a
deterministic evolution.

\section{Discussion}
\begin{figure}[tbh]
\centering \includegraphics[width=0.8\columnwidth]{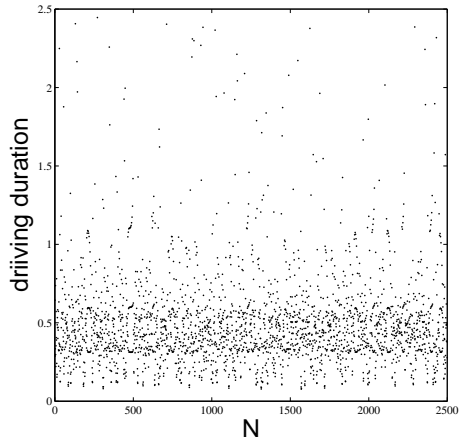}
\caption{Deterministic drive: the duration of the driving episodes.
N represents $N^{th}$ driving event.} \label{tau_ch}
\end{figure}

\begin{figure}[tbh]
\centering \includegraphics[width=0.8\columnwidth]{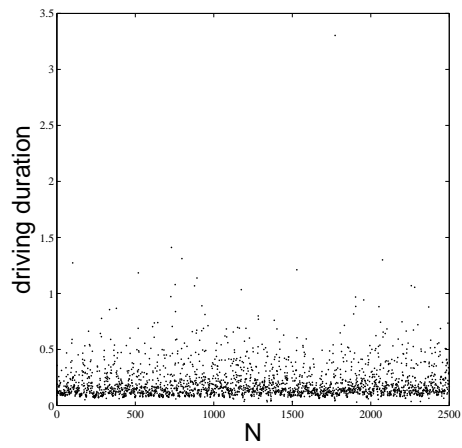}
\caption{Random forcing: the duration of the driving episodes. N
represents $N^{th}$ driving event ($\Delta t ~= ~0.25s$,
$\Omega~=~1.0$, $\gamma~=~0.1$, $s=0.1$ and $a~=~2.5$).}
\label{tau_xi}
\end{figure}

The effect of a forcing on the dynamics can depend on many factors.
The most important are the amplitude and the time for which such a
forcing persists. With Eq. \ref{fliker} the dynamics of the system
in terms of phase space variables is well defined. The only degree
of freedom available for an irregular behavior is through the time
interval in which the system is driven and is not driven. At the
onset of chaos, it can be seen from Fig. \ref{tau_ch} that the
durations for which the driving persists are highly irregular. Also
irregular are the intervals in which the system is not driven. This
intermittent nature of the drive and irregular durations of the
driving events are the only source of irregularities for a dynamics
which is linear in every region of phase space. From Fig.
\ref{tau_xi}, it can be seen that the random forcing applied to the
oscillator also persists for irregular durations. Stochasticity as
such is a source for irregularities in a system. But it may not be
the same as the topological structure of an attractor demands.
There, the intermittent driving and irregular driving durations play
a significant role. It  modifies the irregularities associated with
the  driving suitable for chaotic dynamics in the system. The
durations for which the system is not driven are also irregular in a
similar manner. Thus intermittency is significant to the chaotic
behavior due to its role in synthesizing or seasoning the
complexities that occur in the dynamics.

The damped oscillator, together with its drive, have got another
very fundamental similarity with deterministic chaos. The phase
space of a chaotic system is dense in periodic orbits, or, it
contains infinitely many periodic orbits. None of these periodic
orbits are stable, but they significantly influence the evolution of
a system in the phase space. This aspect is important in considering
the anatomy of the chaotic behavior exhibited by the system under
consideration. Different periodicities arise due to different
numbers of reflections and transmissions in the strip. The drive
acts as the triggering mechanism for reflections or transmissions in
the strip which facilitates smooth transitions between orbits of
various periodicities. This is also the reason why the attractor
structure remains similar under both deterministic and periodic
drivings. Though physically similar, two driving schemes are
completely different in a mathematical sense. Thus intermittently
driven oscillator is an example where determinism can mimic
stochasticity.
\section{Conclusion}
A Damped linear driven oscillator can exhibit chaotic behavior if
the coupling between the drive and the oscillator are applied only
at a narrow region of the phase space. Similar chaotic behavior can
be observed with a purely random forcing. Though physically similar
mechanisms exist for such a behavior, mathematically, they are
different. We hope that this work has revealed yet another relation
between randomness and determinism in nature.

\section{Acknowledgments} One of the author(MPJ) acknowledges the
council of scientific and industrial research (\textbf{CSIR}), New
Delhi for financial support through a senior research fellowship
(\textbf{SRF}). We also thank M. Lakshmanan and S. Rajesh for
fruitful discussions.


\end{document}